\documentstyle[11pt,newpasp,twoside,epsf]{article}
\begin{document}
\title{The Dynamical Interstellar Medium: Insights from Numerical
 Models} \author{Mordecai-Mark Mac Low} \affil{Department of
 Astrophysics, American Museum of Natural History, 79th St. at Central
 Park W., New York, NY, 10024-5192, USA; mordecai@amnh.org}

\begin{abstract}
Numerical models of the dynamical interstellar medium show that
interactions between structures such as supernova remnants and
superbubbles are more important than the structures themselves in
determining the behavior of the ISM.  I review the techniques and
conclusions of recent global models, focussing on what they tell us about
the formation of star-forming giant molecular clouds.
\end{abstract}

\section{Introduction}

The modern picture of the interstellar medium (ISM) as a pervasive
medium containing gas at temperatures ranging from hundreds to
millions of degrees was first hinted at by {\em Copernicus}
observations of the far ultraviolet absorption lines of O~{\sc vi}
(Jenkins \& Meloy 1974, Jenkins 1978).  The ambient interstellar
radiation field is too soft to photoionize O~{\sc vi}, so it must be
thermally ionized by $10^5$~K gas.  However, the cooling curve for
gas with solar abundances peaks around $10^5$~K, implying that this
gas exists in thin, cooling layers at the surface of a reservoir of
$10^6$~K gas.  This led to models of a thermally regulated,
three-phase ISM controlled by the balance between energy input from
supernovae and energy dissipation by radiative cooling (Cox \& Smith
1974; McKee \& Ostriker 1977).  A new model is now beginning to
emerge, in which dynamical compression rather than thermal conduction
determines the thermal state of the gas as well as its structure.

The major question driving this new model is, what determines the star
formation rate?  The old idea that it might be determined by a balance
between supersonic turbulence and self-gravity has come back into
favor recently as the idea of magnetostatic support regulated by
ambipolar diffusion has run into significant problems: magnetic fields
weaker than required (Crutcher 1999); observed cores denser than
predicted (Nakano 1998); and the inability of magnetic fields to
maintain the observed supersonic motions (Mac Low et al.\ 1998, Stone,
Ostriker, \& Gammie 1998).  In this paper I argue that numerical
models point to supernovae as the main driver of that supersonic
turbulence, at least in star-forming regions of the ISM.

As has long been recognized, the odd thing about the observed star
formation rate in normal galaxies is how low it is.  Typical molecular
cloud free-fall times are of order 1~Myr, but there is still
plentiful gas available after $10^4$ times that long.  Another way to
see this is to note that if the $10^{10} M_{\odot}$ of gas observed in
our galaxy were to collapse on timescales approaching the free-fall
time, the star formation rate would be far higher than the observed
$1-2 M_{\odot} \mbox{ yr}^{-1}$.  

This has led to the suggestion that molecular clouds are long-lived
objects, with lifetimes ranging from 30 Myr (Blitz \& Shu 1980) to as
long as 100 Myr (Solomon \& Sanders 1979, Scoville \& Hersh 1979).  In
addition to the argument from the global star formation rate given
above, Blitz \& Shu (1980) also point to the extent of the presence of
molecular clouds behind spiral arm shocks and the stellar ages then
thought to be associated with the clouds as evidence for their
lifetime estimate.  Recently, however, Ballesteros-Paredes, Hartmann,
\& V\'azquez-Semadeni (1999), have proposed much shorter lifetimes of
only 5--7~Myr, based on the lack of post-T~Tauri stars with ages in
the 5--10~Myr range that are clearly associated with molecular clouds.
Individual clouds might be quite short-lived, though cloud-forming
regions might last significantly longer, explaining the observed
spatial relationships with spiral arms.

\section{Simulations}

\subsection{Questions}
Large-scale simulations of the ISM address a number of questions in
ways complementary to other methods.  First, they can begin to reveal
the topology of the different phases of the ISM, distinguishing
between scenarios in which the hot gas is confined in isolated
bubbles, is distributed in an interconnected tunnel network, or forms
a uniform sea surrounding isolated clouds of warm and cold gas.  The
distribution of cold, neutral gas determines whether ionizing
radiation from OB associations in the galactic plane can fully account
for the ionization of the Reynolds layer far above it.

Second, the production of star-forming regions appears to be largely
determined by large-scale flows in the ISM, influenced by the
gravitational potential.  Large-scale simulations can thus yield
information on the formation and lifetimes of molecular clouds.
Related issues can also be addressed, for example, whether
supernovae and superbubbles are more important in triggering nearby
star formation or in driving turbulence that supports the ISM against
gravitational collapse and star formation.

Finally, two other major issues that I will not focus on to the same
extent are the production of galactic magnetic fields in a turbulent
dynamo, and the vertical flow of the ISM in galactic fountains and
winds. 

\subsection{Objects vs.\ Interactions}

Traditionally, numerical simulations of the ISM focussed on specific
objects, such as supernova remnants, stellar wind bubbles,
superbubbles, shocked clouds, or bow shocks produced by fast-moving
stars or stellar jets.  Such simulations have proved useful, but are
limited by the lack of interactions with surrounding structures.

Observations of the ISM are structured at all scales.  A frequent
problem encountered by observers is distinguishing objects such as
superbubbles from the structured background.  The background structure
is often as important in defining the final nature of an object as
more traditional considerations such as the central energy source or
cooling physics.  The interactions between ``objects'' such as
supernova remnants or superbubbles becomes as important as the objects
themselves.  The random supersonic flows generated by these
interactions in turn determines whether regions will be supported
against gravitational collapse, or be compressed and cooled, starting
a process of collapse leading to molecular cloud and star formation.

\subsection{Models}

In the last five years, a significant number of global models have
been computed.  They are distinguished by the different physical
processes included, and by the different numerical techniques used.  

The most important distinction appears to be the driving mechanism for
the modeled interstellar turbulence.  The most abstract model for the
driving is simply a uniform, Gaussian field, which has both the
advantage and disadvantage of being very general, but not reproducing
the specifics of particular real drivers.  Three real drivers have
also been used in various models: stellar ionization heating,
rotational shear, and supernovae.  The last appears to be the most
realistic, although the other two add important physics when used in
addition to the supernovae.

The dimensionality and coverage of the simulation are also very
important.  Two-dimensional simulations overstate the difficulty of
hot gas escaping from the plane and cannot be used fruitfully to
explore questions of magnetic field generation, but have yielded
fundamental insights into the turbulent ISM.  Three-dimensional
models have been done in both local and global modes, either focussing
on a particular region of a disk (typically around 1 kpc$^2$) or
attempting to encompass the entire disk.

Three main technical issues differentiate the models.  First is the
numerical diffusivity of the method used.  Smoothed particle
hydrodynamics (SPH) tends to be the most diffusive method due to the
need to average over a fairly large number of particles to derive the
flow quantities.  Spectral methods are excellent for smooth flows, but
handle shocks and other abrupt features in flows poorly.  Grid-based
methods have a numerical diffusivity determined by a combination of
the order of the method, the advection technique used
(e.g. second-order Van Leer, piecewise parabolic method (PPM), or
total variation diminishing (TVD) methods), and the technique used for
handling shocks.

The second main issue is the shock handling technique itself.
Artificial viscosity can be used with particle, spectral, and grid
methods to resolve shocks over several resolution lengths.  Exact or
linearized Riemann solvers resolve shocks more accurately and in fewer
zones.  They compute the flow at the edge of each zone by solving for
the combination of shocks and compression or rarefaction waves that
results from the breakup of the discontinuity at each zone boundary.

The third issue is the structure of the grid in those methods using
one.  The simplest grids are fixed, Eulerian grids, with periodic
boundary conditions in the case of local simulations.  Ratioed fixed
grids that focus resolution toward a region of interest such as the
galactic plane or the galactic center are the next step.  SPH models
effectively use an unstructured Lagrangian grid that follows the flow,
with the advantage of high resolution in dense regions, but the
corresponding disadvantage of low resolution in more rarefied regions.
Adaptive mesh refinement focusses resolution into regions of
interest regardless of density, at the cost of significant additional
programming complexity.

\section{Results}

The outlines of a coherent picture of the interstellar medium is
emerging from the numerical models, one that places a greater emphasis
on dynamical interactions and less on equilibrium structures than
previous analytically constructed pictures.

In the absence of supernova heating and the presence of normal
interstellar cooling, the gas cools rapidly, forming dense clouds
scattered through the disk of the galaxy that interact with each
other, surrounded by a thicker disk of warm $10^4$~K gas, but without
hot, $10^6$~K, X-ray emitting gas. An example of this solution is seen
in the global SPH computations by Gerritsen \& Icke (1997).

The global, two-dimensional PPM computations of Wada \& Norman (1999)
add the element of driving from galactic shear. Starting with a Toomre
stable disk, they find that the gas rapidly cools and collapses into a
filamentary network of clouds that are cold enough to be Toomre
unstable. Although occasional cloud collisions generate small amounts
of $10^5$~K gas, the interesting conclusion can be drawn from these
computations that the Toomre stability criterion must be applied to a
medium whose velocity dispersion is maintained by some mechanism other
than ionization heating.

At the same time, it has become clear that dense interstellar clouds
can be formed by turbulent compression.  The local, 2D, MHD, spectral
models of Passot, V\'azquez-Semadeni, \& Pouquet (1995) show that
this can be a dominant mechanism for structure formation in the warm
and cold ISM even in the presence of magnetic fields.  Hennebelle \&
P\'erault (1999, 2000) have explored the interplay of compression and
thermal condensation due to conduction in detail with 1D, adaptive
mesh models with and without magnetic fields, showing that compression
will cause collapse on far shorter timescales than thermal
condensation under typical ISM conditions.  As thermal condensation is
the dominant mechanism for cold cloud formation in analytic
equilibrium models such as those of McKee \& Ostriker (1977), this is
an important conclusion.

Using supernovae as the main driving mechanism produces hot gas with
moderate filling factor in the disk of galaxies with parameters
similar to those of the Milky Way, and much larger filling factor
above the plane.  Local, 2D models done with a second-order, Van Leer
algorithm on a ratioed grid by Rosen \& Bregman (1995) established
that such driving alone could lead to the formation of a network of
cold clouds and hot gas, along with a flow of hot gas out of the disk.
Use of adaptive mesh refinement with a Riemann solver allowed Avillez
(1999) to perform 3D models extending 10~kpc into the upper halo.  He
finds that hot gas is not trapped in the disk as occurred in the 2D
models.  Ballesteros-Paredes, Avillez, \& Mac Low (2000) show that
instead, isolated, cold, dense clouds with masses and sizes typical of
giant molecular clouds form in and close to the disk.  Column density
distributions of cool and cold gas from these models strongly resemble
observations of H~{\sc i}, for example by Kim, Stavely-Smith \&
Bessell (1999).

These models either lack self-gravity, or, in the case of Passot et
al.\ (1995) lack the resolution for self-gravity to become
significant.  However, isothermal, uniformly-driven models provide
some insight into whether supernova-driven turbulence could support
the gas against gravitational collapse.  Klessen, Heitsch, \& Mac Low
(2000) show that supersonic turbulence can globally support a
thermally Jeans unstable region, but local collapse will still occur
due to the density enhancements caused by shock waves in the
supersonic turbulence, except in the case of extremely strong
turbulence.  In fact, it appears that a signpost of global turbulent
support under molecular cloud conditions is isolated star formation,
while lack of turbulent support leads to efficient runaway star
formation (Klessen, Burkert, \& Bate 1998, Klessen \& Burkert 2000a,
2000b).  Heitsch, Mac Low, \& Klessen (2000) show that magnetic fields
add small-scale structure, and slow local collapse, but do not prevent
it.

Applying the insights gained from these models to the interstellar
case, it appears that the supernova-driven turbulence may naturally
create regions that are gravitationally-bound, but then may drive
enough turbulence within them to reduce their star-formation rate to
the low levels observed in typical molecular clouds, probably with the
help of interstellar magnetic fields.  Observations of molecular
clouds appear to show that they have self-similar structure to scales
of as much as tens of pc (e.g.\ Stutzki et al. 1998), which Mac Low \&
Ossenkopf (2000) find to be most readily explained by driving of the
turbulence from even larger scales, again consistent with large-scale
driving by supernovae.

Korpi et al.\ (1999) use a 3D, fixed grid, local, MHD code to model
the evolution of the magnetic field in shearing, supernova-driven
turbulence.  Korpi (1999) measures the stress terms in the flow to
find preliminary evidence for the operation of a dynamo, with roughly
the strength predicted by Ferri\`ere (1996, 1998) from semi-analytic
models of superbubble expansion. 

All the supernova-driven models find a flow of hot gas from the plane
into the halo.  Avillez \& Mac Low (2000) point out that much of this
gas does not flow through chimneys directly into the halo, but rather
rises buoyantly into the Lockman and Reynolds layers, not necessarily
reaching the hot halo.  A true fountain flow only appears above 1 kpc
(Avillez 2000).  One consequence of this is that even fairly active
normal galaxies, such as the Milky Way or NGC~891, do not drive strong
winds into the surrounding intergalactic medium.  Kinetic energy
feedback from galaxy formation will only come from starburst galaxies
(Mac Low \& Ferrara 1999, Silich \& Tenorio-Tagle 1998, Suchkov et
al.\ 1994, 1996).

\acknowledgements I thank the conference organizers for partial
support of my attendence.  This work was partially supported by the US
NSF under CAREER grant AST 99-85392, and made use of the NASA
Astrophysical Data System Abstract Service.

\end{document}